\documentclass[aps,prd,amsmath,superscriptaddress,twocolumn,nofootinbib]{revtex4-1}  % for review and submission
\usepackage{graphicx}  % needed for figures
\usepackage{dcolumn}   % needed for some tables
\usepackage{bm}        % for math
\usepackage{verbatim}   % for math
\usepackage{amssymb}   % for math
\usepackage{amsmath}
\usepackage{txfonts}
\usepackage{latexsym}
\usepackage{amsbsy}
\usepackage{mathrsfs}
\usepackage{color}
\usepackage{color,bm}

\begin{document}

\title{Origin of the fundamental plane of elliptical galaxies in the Coma Cluster without fine-tuning}
\author{Mu-Chen Chiu}
\email[]{chiumuchen@gmail.com}
\affiliation{
Shanghai Key Lab for Astrophysics, Shanghai Normal University, Shanghai 200234, China
   }
\author{Chung-Ming Ko}
\email[]{cmko@astro.ncu.edu.tw}
\affiliation{
Institute of Astronomy, Department of Physics and Center for Complex Systems, National Central University, Zongli Dist., Taoyuan City, Taiwan (R.O.C.)
   }

   \author{Chenggang Shu}
   \email[]{cgshu@shao.ac.cn}
\affiliation{
Shanghai Key Lab for Astrophysics, Shanghai Normal University, Shanghai 200234, China
   }
\date{\today}
\begin{abstract}
After thirty years of the discovery of the fundamental plane, explanations to the tilt of the fundamental plane with respect to the virial plane still suffer from the need of fine-tuning. In this paper, we try to explore the origin of this tilt from the perspective of modified Newtonian dynamics (MOND) by applying the 16 Coma galaxies available in~\citet{Thomas_etal11}. Based on the mass models that can reproduce de Vaucouleurs' law closely, we find that the tilt of the traditional fundamental plane is naturally explained by the simple form of the MONDian interpolating function, if we assume a well motivated choice of anisotropic velocity distribution, and adopt the Kroupa or Salpeter stellar mass-to-light ratio. Our analysis does not necessarily rule out a varying stellar mass-to-light ratio.
 \end{abstract}

\maketitle

\section{Introduction}

The first scaling law of elliptical galaxies, the Faber-Jackson relation, was found by~\citep{faber&jackson76} in 1976, and it was later incorporated into a fundamental plane (FP) between the size, surface brightness, and velocity dispersion of elliptical galaxies~\citep{Djorgovski&Davis87, Dressler_etal87,Jorgensen_etal96}.
However, this empirical FP does not comply with the dimensional-analysis of the virial theorem in the standard Newtonian dynamics while a constant total mass-to-light ratio ($M/L$) is assumed~\citep[e.g.][]{Gerhard_etal01,Treu_etal05}. This deviation of the observed FP to the theoretical prediction is called the tilt of the FP. The tilt is usually attributed to a varying total $M/L$ that increase with mass. In the past decade, thanks to the independent, accurate measurement of total mass by stellar dynamics~\citep{Cappellari06,Thomas_etal11,Cappellari13} or strong lensing~\citep{Bolton_etal07,Bolton_etal08,Auger_etal10}, a genuine variation of total $M/L$  has been confirmed. Not only the total $M/L$ derived from these two different techniques are consistent, but also both techniques show that other than the FP, there is a less scattering total mass plane (MP) that agrees well with the prediction of the virial theorem~\citep{Bolton_etal08,Thomas_etal11,Cappellari13}. Based on this independent estimation of total mass, it has also been shown that the total $M/L$ is highly correlated to the residuals of the FP, so a variation of $M/L$ among galaxies is responsible for the tilt of the FP~\citep{Bolton_etal07}.

Despite of the clear evidences of the systematic variation of total $M/L$ among galaxies, the origin of this variation, so the tilt of the FP, is an ongoing debate~\citep{Desmond&Wechsler16,D'Onofrio_etal16}.
 Explanations include structural non-homology in elliptical galaxies~\citep{Trujillo_etal04}, variations in the distribution of dark matter~\citep{Renzini&Ciotti93}, and the systematic variation of stellar population~\citep{Forbes98}. Of these, non-homology not only requires fine-tuning between stellar $M/L$ and Sersic index, but also just plays a minor role in the tilt of the FP at best~\citep{Bolton_etal08,Cappellari13}. On the other hand, the total $M/L$ used to derive the MP in~\cite{Thomas_etal11} obviously differs from the stellar $M/L$ estimated from the population synthesis, which implies that the tilt of the FP is not a pure stellar population effect. Instead, either a non-universal stellar initial mass function (IMF) or variations of dark matter fraction among systems shall account for, at least partly, the tilt of the FP. However, the explanations from IMF and dark matter both require strong fine-tuning~\citep{Ciotti96}.

 Since the nature of dark matter can contribute to the tilt of the FP, it is only one step forward to ask whether or not the tilt of the FP can also be attributed to other alternatives to dark matter. Modified Newtonian dynamics (MOND), as an alternative paradigm to dark matter~\citep{Milgrom83a}, is very successful at galactic systems, not only the overwhelmingly convincing baryonic Tully-Fisher relation~\citep{McGaugh05,McGaugh11,McGaugh12} and rotation-curve analysis in spiral galaxies~\citep{Sanders&McGaugh02,Famaey&McGaugh12}, but also analysis of galactic strong lensing~\citep{Chiu_etal06,Chiu_etal11,Zhao_etal06,Tian_etal13}, galaxy-galaxy lensing~\citep{Milgrom13}, dynamical analysis of planetary nebulae~\citep{Tian&Ko16}, the velocity dispersions of the dwarf satellites of Andromeda~\citep{McGaugh&Milgrom13a,McGaugh&Milgrom13b,Pawlowski&McGaugh14}, gas ring in elliptical galaxies~\citep{Milgrom12}, and newly found relation between the baryonic and dynamical central surface densities of disc galaxies~\citep{Lelli_etal16,Milgrom16}, and radial acceleration in rotating galaxies~\citep{McGaugh16}. However, although the Tully-Fisher relation of spiral galaxies is naturally explained in MOND~\citep{Milgrom83b}, the origin of the FP of elliptical galaxies in MOND is much less clear. Several papers have attempted to explain the FP in MOND~\citep{sanders00,sanders&land08,Cardone_etal11}, but none of them have explicitly applied observational data to compare exponents of the observed MOND `fundamental plane' with their theoretical counterpart. To the best of our knowledge, this paper is the first attempt to do so.

 In this paper, we apply $16$ Coma elliptical galaxies in~\citet{Thomas_etal11} to construct the FP and MP in MOND, where stellar mass is tantamount to total mass, because the stellar $M/L$ estimated from stellar population synthesis is available for every member of these galaxies. In addition, since some of these galaxies have more than $50\%$ of dark matter within one effective radius, $\rm{R}_{\rm{e}}$~\citep{Thomas_etal11}, we expect a substantial MONDian effect will be detected in these data.

The structure of the paper is organized as below. In Sec.~\ref{II}, we discuss the dynamics of elliptical galaxies in MOND, and show how to obtain the theoretical MOND FP in the following section. We then construct the observational MOND FPs from the $16$ Coma elliptical galaxies, and compare them with their theoretical counterparts in Sec.~\ref{IV}. Finally, we give a brief discussion and conclusion in Sec.~\ref{V}.

\section{Dynamics of elliptical galaxies in MOND}\label{II}
For spherical systems, the Jeans equation in the context of MOND only differs from its counterpart in the Newtonian dynamics in the radial gravitational force, $g=\tilde{\nu}(x_N)g_N$~\citep{sanders00,Cardone_etal11,chae&gong15,Tian&Ko16}, where $x_N$ is the ratio between the Newtonian gravity, $g_N$, and the MOND acceleration constant, $a_0$ ($=1.2\times10^{-8}\rm{cm\,s}^{-2}$), and where $\tilde{\nu}$ is the interpolating function of MOND~\citep[e.g.][]{Milgrom83a,Chiu_etal11,Ko16}. Following exactly the same way as in the Newtonian dynamics~\citep[e.g.][]{binney&mamon82,agnello_etal14}, we can solve for
the aperture velocity dispersion, $\sigma_{\rm{ap}}$, from the modified Jeans equation in MOND with certain matter density and anisotropy profile of velocity distribution, $\beta$. For the matter density, we assume that mass always follows light, so the Hernquist model~\citep{hernquist90}, which approximates de Vaucouleurs' law closely, shall describe all mass in elliptical galaxies. Accordingly, we obtain
\begin{equation}
    \sigma_{\rm{ap}}^2=\frac{G \mathcal{M}}{{R}_h} \mathcal{H}(\tilde{R})
                 {\;}\Gamma^{M}_{h}(\tilde{\it{R}},\beta,\Upsilon {\,} {I}_e)
	\label{eq:sig_spl}
\end{equation}
where $G$ is Newton's constant, $\mathcal{M}$ is total mass of galaxy, $\Upsilon$ is stellar mass-to-light ratio, $I_e$ is surface brightness within the effective radius, and $R_h$, a characteristic length in the Hernquist model, is approximately related to the effective radius by $0.551$$R_e$.  In Eq.~(\ref{eq:sig_spl}), we also denote $\mathcal{H}(\tilde{R})$ as a ratio between total luminosity and surface brightness within $\tilde{R}$, which itself is a radius normalised to $R_h$, and use
\begin{eqnarray}
\Gamma^{M}_{h}= 2 \int^{\tilde{\it{R}}}_0
                    \tilde{\it{R}}' d \tilde{\it{R}}' \int^{\infty}_{\tilde{\it{R}}'}\frac{\tilde{\it{r}} d \tilde{\it{r}}}{\sqrt{\tilde{\it{r}}^2-\tilde{\it{R}}'^2}}
                    \Big(1-\beta(\tilde{\it{r}})\frac{\tilde{{\it{R}}}'^2}{\tilde{\it{r}}^2}\Big) \nonumber\\
                        \int^{\infty}_{\tilde{\it{r}}} \frac{\tilde{\nu}(\Upsilon {\,} I_e,\tilde{\it{r}}')}{\tilde{\it{r}}'(\tilde{\it{r}}'+1)^5}
                       \exp\Big(\int^{\tilde{\it{r}}'}_{\tilde{\it{r}}} \frac{2\beta}{{\tilde{\it{r}}''}}{d \tilde{\it{r}}''} \Big) {d \tilde{\it{r}}'}
	\label{eq:Gamma}
\end{eqnarray}
to describe the component that will be influenced by MOND. For any fixed aperture radius, $\Gamma^{M}_{h}$ will in general depend on anisotropy, surface brightness within the effective radius, mass-to-light ratio, and the MONDian interpolating function. On the contrary, when the MONDian effect vanishes, $\Gamma^{M}_{h}$ will only depend on $\beta$. We denote this special case of $\Gamma^{M}_{h}$ as $\Gamma^N_{h}(\tilde{R},\beta)$. According to the notations above, Eq.~(\ref{eq:sig_spl}) can be rewritten as
\begin{equation}
    \sigma_{\rm{ap}}^2=2 \pi G \times \eta(\tilde{R},\beta) \times \Upsilon \times R_e \times I_e \times \Xi^{M}(\tilde{\it{R}},\beta,\Upsilon {\,} I_e),
	\label{eq:sig_ap}
\end{equation}
where
\begin{equation}
    \eta(\tilde{R},\beta)=R_h/R_e \times \mathcal{H}(\tilde{R}) \Gamma^N_{h}(\tilde{R},\beta)
	\label{eq:eta}
\end{equation}
is defined as a structural parameter that is solely decided by mass model and anisotropy, and
\begin{equation}
    \Xi^{M}=\Gamma^M_{h}/\Gamma^N_{h}
	\label{eq:Xi}
\end{equation}
is a MONDian parameter that describes how much $\sigma_{\rm{ap}}$ will deviate from its Newtonian counterpart. Basically, Eq.~(\ref{eq:sig_ap}) plays a role similar to the virial theorem in the traditional analysis of the FP.

 We would like to emphasize that although the derivation of Eq.~(\ref{eq:sig_ap}) is based on the Hernquist model, we can easily extend it to other mass models, and will obtain similar results. For example, for the Jaffe model~\citep{jaffe83}, which is another analytical mass profile that successfully reproduces de Vaucouleurs' law, we can obtain an equation similar to Eq.~(\ref{eq:sig_ap}) except for replacing $\Gamma^M_{h}$ in the structural parameter, $\eta$, and the MONDian parameter, $\Xi^{M}$, with
 \begin{eqnarray}
\Gamma^{M}_{j}= 2 \int^{\tilde{\it{R}_j}}_0
                    \tilde{\it{R}_j}' d \tilde{\it{R}_j}' \int^{\infty}_{\tilde{\it{R}_j}'}\frac{\tilde{\it{r}_j} d \tilde{\it{r}_j}}{\sqrt{\tilde{\it{r}_j}^2-\tilde{\it{R}_j}'^2}}
                    \Big(1-\beta(\tilde{\it{r}_j})\frac{\tilde{{\it{R}_j}}'^2}{\tilde{\it{r}_j}^2}\Big) \nonumber\\
                    \int^{\infty}_{\tilde{\it{r}_j}} \frac{\tilde{\nu}(\Upsilon {\,} I_e,\tilde{\it{r}_j}')}{\tilde{\it{r}_j}'^3(\tilde{\it{r}_j}'+1)^3}
                       \exp\Big(\int^{\tilde{\it{r}_j}'}_{\tilde{\it{r}_j}} \frac{2\beta}{{\tilde{\it{r}_j}''}}{d \tilde{\it{r}_j}''} \Big) {d \tilde{\it{r}_j}'},
	\label{eq:Gamma_j}
\end{eqnarray}
 where the radii are normalised to the characteristic length in the Jaffe model, $R_j=1.3106 R_e$, and for substituting $R_h$ and $\mathcal{H}(\tilde{R})$ in the structural parameter with $R_j$ and $\mathcal{H}(\tilde{R}_j)$.

%As shown above, the influence of the MONDian effects, which is channeled solely via $\Xi^{M}$, depend on four factors: $\beta$, $\rm{I}_{\rm{e}}$, $\Upsilon$, and $\tilde{\nu}$. Of these, only $\rm{I}_{\rm{e}}$ can be directly determined by observations. Before exploring whether or not it requires fine-tuning of the other three factors to match the observed MFP with the theoretical prediction, we will discuss how these factors exert influence on $\Xi^{M}$ first.

\section{The MONDian Fundamental Plane}

\subsection{Anisotropy, Mass-to-light ratios, and MONDian effects}\label{sec:3}
\begin{figure}
\includegraphics[width=\columnwidth]{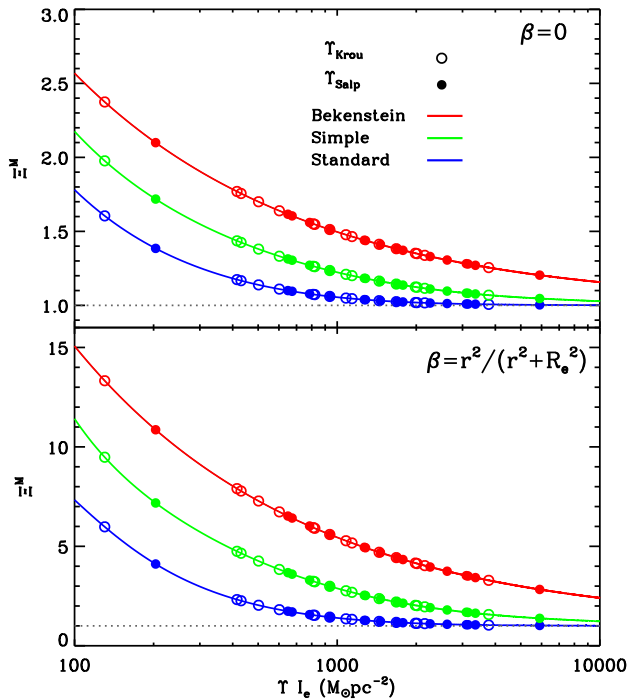}
\caption{\label{fig1} The MONDian parameter $\Xi^{M}$ defined in Eq.~(\ref{eq:Xi}) is plotted against $\Upsilon {I}_e$ for various MONDian interpolating functions and anisotropies. We also plot $\Xi^{M}$ of the 16 Coma galaxies with two different universal stellar $M/L$: $\Upsilon_{\rm{Krou}}$ and $\Upsilon_{\rm{Salp}}$. When the MONDian effects vanish, $\Xi^{M}$ will always equal one (dotted line).}
\end{figure}

In this paper, we only consider three most well-known forms of the MONDian interpolating function, that is, the standard, simple, and Bekenstein forms~\citep[e.g.][]{Famaey&McGaugh12}. As shown in Fig.~\ref{fig1}, under the same conditions, the Bekenstein form will yield the strongest MONDian effect to boost the velocity of dispersion; on the contrary, the standard form will lead to the weakest. Figure~\ref{fig1} also shows that for a given surface brightness, the larger the stellar $M/L$, the weaker the MONDian effect; on the other hand, like spiral galaxies~\citep{mcgaugh&deblock}, for a fixed stellar $M/L$, the lower surface brightness will have stronger MONDian effects. In addition, Fig.~\ref{fig1} demonstrates the values of $\Xi^{M}$ of the 16 Coma galaxies considered in this paper for two different types of $\Upsilon$: the Kroupa ($\Upsilon_{\rm{Krou}}$) and Salpeter ($\Upsilon_{\rm{Salp}}$) mass-to-light ratio in~\citet{Thomas_etal11}. The values of $I_e\Upsilon_{\rm{Krou}}$ and $I_{e}\Upsilon_{\rm{Salp}}$ of these 16 galaxies range from around $100$ to $6000$ $\rm{M}_{\odot} \rm{pc}^{-2}$. Finally, we compare the influences of anisotropy on $\Xi^{M}$. We consider anisotropy profile $\beta=r^2/(r^2+r_a^2)$~\citep{vanAlbada82,sanders00,Milgrom&sanders03,agnello_etal14}, which fits well the velocity distributions of the systems formed by dissipationless collapse~\citep{vanAlbada82,sanders00}, but set $r_a=R_e$ to avoid an ad hoc free parameter, and obtain
\begin{equation}
\beta_e(r)=\frac{r^2}{(r^2+R_{e}^2)}.
\label{beta_e}
\end{equation}
To check the sensitivity of $r_a$, we also consider $r_a=3 R_e$, which is used by~\citet{Milgrom&sanders03} to fit the line-of-sight velocity dispersion and the planetary nebulae observations. This alternation will only cause less than $5\%$ changes of $\Xi^M$.
Based on $\beta_e$, we find the stark differences of the influences of anisotropy profile between the Newtonian dynamics and MOND, which have never been noticed before. In the Newtonian dynamics, the contribution of $\beta_e$ to the boost of $\sigma_{\rm{ap}}$ is a constant, and only $10\%$ more than its counterpart in the case of isotropy. However, in MOND, the impact of introducing $\beta_e$ into Eq.~(\ref{eq:Gamma}) is not only much larger, but also not a constant: it depends on $\Upsilon$, $I_e$, and $\tilde{\nu}$.

\subsection{The tilt of the Fundamental Plane}
While we assume a universal stellar $M/L$ in elliptical galaxies, the existence of a theoretical MOND FP between $R_e$, ${\sigma_{\rm{ap}}}$, and $I_e{\Xi^{M}}$ can be easily recognized from Eq.~(\ref{eq:sig_ap}), which leads to
\begin{equation}
    \log{\frac{R_e}{\rm{kpc}}}= a \log{\frac{\sigma_{\rm{ap}}}{\rm{km}\;\rm{s}^{-1}}} + b \log{\left(\frac{ I_e}{\rm{L}_{\odot}\;\rm{pc}^{-2}}{\Xi^{M}}\right)}+c,
	\label{eq:MFP}
\end{equation}
with $a=2$, $b=-1$, and
\begin{equation}
    c=  - \log{\frac{\langle\Upsilon\rangle}{\rm{M}_{\odot}/\rm{L}_{\odot}}}- \log{\left(2\pi\frac{ G}{(\rm{km}\;\rm{s}^{-1})^2\;\rm{kpc}\;{\rm{M}_{\odot}^{-1}}}\eta\right)}.
	\label{eq:c_MFP}
\end{equation}
Here, we take the mean of $\Upsilon$, $\langle\Upsilon\rangle$, to guarantee that $c$ is a constant because even for a universal stellar $M/L$, there must exist some degree of random variations of $\Upsilon$ for individual galaxies.
Equation~(\ref{eq:MFP}) is identical to its Newtonian counterpart except for the additional parameter of the MONDian effects, $\Xi^{M}$, which implicitly includes the MOND acceleration constant, $a_0=1.2\times10^{-8}\rm{cm\,s}^{-2}$. Unlike the Newtonian FP derived from the virial theorem, however, we do not use a free parameter (i.e. a structure coefficient) to ambiguously represent galactic structure~\citep[e.g.][]{Bolton_etal07,Thomas_etal11,Cappellari13}; instead, the influence of galactic structure on the MOND FP is analytically decided by mass models of galaxies. Since we assume a Hernquist model throughout our derivation, according to Eq.~(\ref{eq:eta}), the structure parameter, $\eta$, will only depend on $\beta$ for any fixed radius, $\tilde{R}$. This fact also implies that homology of galaxies (galactic structure is independent of the galaxy mass) is implicitly assumed in Eq.~(\ref{eq:MFP}).

Following~\citet{Bolton_etal07}, we might also incorporate variations of stellar $M/L$ among galaxies into the MOND FP by
combining $\Upsilon$ with $I_e$ in Eq.~(\ref{eq:MFP}), so that surface brightness will be replaced with surface matter density, and the MOND FP will become the MOND `mass plane',
\begin{equation}
    \log{\frac{{R}_e}{\rm{kpc}}}= a \log{\frac{\sigma_{\rm{ap}}}{\rm{km}\;\rm{s}^{-1}}} + b \log{\left(\frac{ {\Sigma}_e}{\rm{M}_{\odot}\;\rm{pc}^{-2}}{\Xi^{M}}\right)}+c,
	\label{eq:MMP}
\end{equation} where ${\Sigma}_e$ is the surface matter density within the effective radius, and the coefficients $a$ and $b$ remain the same as in Eq.~(\ref{eq:MFP}), but
\begin{equation}
    c=  - \log {\left(2\pi\frac{ G}{(\rm{km}\;\rm{s}^{-1})^2\;\rm{kpc}\;{\rm{M}_{\odot}^{-1}}}\eta\right)}.
	\label{eq:c_MMP}
\end{equation}
This is the first time that the MOND FP and MOND MP are written down explicitly and lucidly. According to Eqs~(\ref{eq:MFP})-(\ref{eq:c_MMP}), the tilt of the FP is attributed to a single parameter, $\Xi^{M}$.

Although ideally the MOND FP and MOND MP are identical for a universal stellar $M/L$, in practice, these two planes are not exactly the same. Firstly, we need $\langle\Upsilon\rangle$ in Eq.~(\ref{eq:c_MFP}), but individual $\Upsilon$ to calculate $\Sigma_e$ in Eq.~(\ref{eq:MMP}).  Secondly, even if there is no random variation around $\langle\Upsilon\rangle$, the MOND MP will still differ from the MOND FP in the exponent of $\Upsilon$, which together with  $I_e$ is implicitly bound up with the exponent of $\Sigma_e$ in the MOND MP, but is always equal to $-1$ in the MOND FP. These subtle differences are non-trivial when we try to find and compare best-fitting coefficients of the MOND FP and the MOND MP.

\section{Data analysis}\label{IV}

\begin{figure*}
\includegraphics[width=1.0\textwidth]{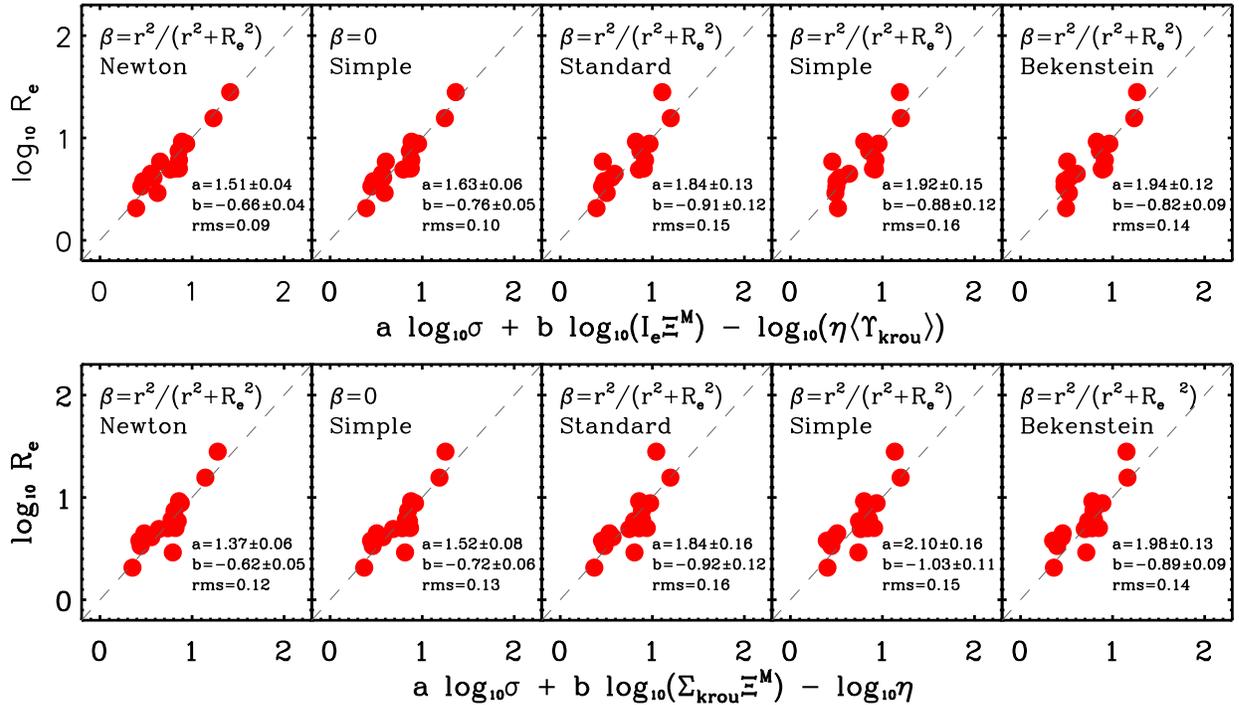}
\caption{\label{fig2}The MOND FP (upper panels) and MOND MP (lower panels) constructed from the 16 Comma galaxies for $\Upsilon_{\rm{Krou}}$, various MONDian interpolating functions and anisotropies. Both MOND FP and MOND MP agree well with their theoretical predictions, Eq.~(\ref{eq:MFP}) and Eq.~(\ref{eq:MMP}), when the acceleration constant,  $a_0=1.2\times10^{-8}\rm{cm\,s}^{-2}$, the simple form of MONDian interpolation function~\citep[e.g.][]{Famaey&McGaugh12}, and the anisotropy profile, Eq.~(\ref{beta_e}), are assumed.}
\end{figure*}

Similar to \citet{Bolton_etal07} and \citet{Thomas_etal11}, we perform regression analysis of the 16 Coma galaxies in order to find the best-fitting coefficients of the observed MOND FP and MOND MP for various $\Upsilon$, $\beta$, and $\tilde{\nu}$ outlined above. The error bars of these coefficients are estimated from one standard deviation of the mean of bootstrap resampling. Unlike the traditional regression analysis of the FP~\citep[e.g.][]{Bolton_etal07,Thomas_etal11,Cappellari13}, however, the intercepts of the observed MOND FP and MOND MP in our analysis are not decided by the best-fitting value, but calculated from Eq.~(\ref{eq:c_MFP}) and Eq.~(\ref{eq:c_MMP}) accordingly, where $\Upsilon$ in Eq.~(\ref{eq:c_MFP}) is taken from the mean value of $\Upsilon_{\rm{Salp}}$ or $\Upsilon_{\rm{Krou}}$ of the 16 Coma galaxies in~\citet{Thomas_etal11}. By adopting the no-intercept regression analysis, we will have fewer free parameters to fit.

%Although the slope found by a no-intercept regression might be biased toward the prediction of a chosen model, as we are going to show below, in our case this type of regression analysis has an advantage to avoid finding results that are totally unphysical.
Firstly, we study the FP and MP of the 16 galaxies without any MONDian effect (left end of Fig.~\ref{fig2}). Since the contributions of $\beta_e$ to the these planes are independent of $\sigma_{\rm{ap}}$, $R_e$, and $I_e$, we do not expect any obvious deviation of our results from previous studies. The consistency of our fit with that of~\citep{bernardi_etal03,Thomas_etal11} within the statistical uncertainties reinforces this expectation.

Among various assumptions of the dynamics of elliptical galaxies in MOND, we find that the assumption of isotropic velocity dispersion of galaxies ($\beta=0$) will lead to great difficulties in removing the tilts of the empirical MOND FP and MOND MP with respect to their theoretical counterparts (the left end of Fig.~\ref{fig2}). Indeed, we can hardly improve the tilts in MOND for $\beta=0$ whatever the interpolating functions we choose. In contrast, while taking the anisotropy profile $\beta_e$, we find that the observed MOND FPs and MOND MPs both agree well with their theoretical counterparts along with a universal stellar $M/L$ and simple form of $\tilde{\nu}$.
Under these conditions, the best-fitting coefficients of the MOND FP for $\Upsilon_{\rm{Krou}}$ are $a=1.92\pm0.15$ and $b=-0.88\pm0.12$ with ${rms}=0.16$, and that of the MOND MP are $a=2.10\pm0.16$ and $b=-1.03\pm0.11$ with ${rms}=0.15$ (Figure~\ref{fig2}). Similarly, if we replace $\Upsilon_{\rm{Krou}}$ by $\Upsilon_{\rm{Salp}}$ but leave the other conditions unchanged, we will obtain the best-fitting coefficients, $a=1.96\pm0.13$ and $b=-0.87\pm0.11$, of the MOND FP as well as $a=2.04\pm0.16$ and $b=-0.96\pm0.10$ for the MOND MP;  both planes have ${rms}=0.15$. It is worth noting that we can hardly detect any change of the slopes if we replace ${R}_e$ with $3 R_e$ in $\beta_e$. In addition, the larger rms errors of the MOND FPs and MOND MPs of $\beta_e$, compared to that of $\beta=0$, might be largely attributed to the fact that although $\beta_e$ contributes a lot to the slopes of MOND FPs and MOND MPs, it is at best an averaged approximation to an ensemble of real anisotropies of the 16 coma galaxies. Unlike the simple form, although the standard form with $\beta_e$ yields a very good value of the coefficient $b$, its another best-fitting coefficient $a$ is not so good; similarly, in the case of the Bekenstein form, the best-fitting coefficient $b$ is not as good as $a$.

\begin{figure*}
\includegraphics[width=1.0\textwidth]{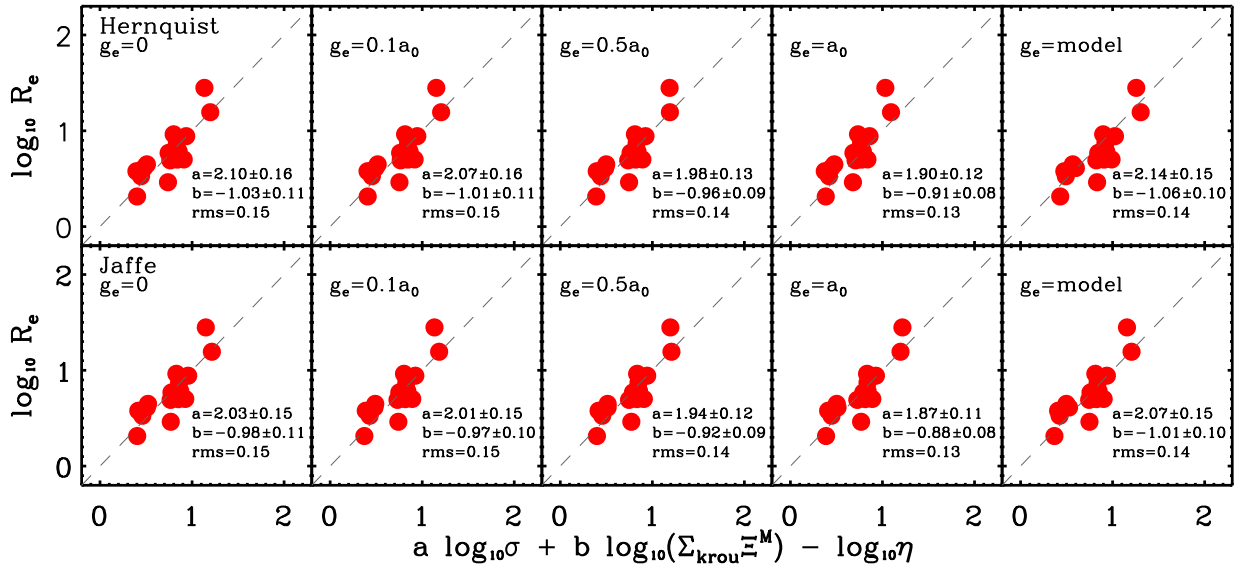}
\caption{\label{fig3}The MOND MP constructed from the 16 Comma galaxies under various values of external field for both the Hernquist and Jaffe models. The model of the external field on the right end side of the figure is explained in the text. The acceleration constant, $a_0=1.2\times10^{-8}\rm{cm\,s}^{-2}$, the simple form of MONDian interpolation function~\citep[e.g.][]{Famaey&McGaugh12}, and the anisotropy profile, Eq.~(\ref{beta_e}), are assumed}
\end{figure*}

In clusters of galaxies, the MONDian effect in elliptical galaxies might be affected by the external fields of the host clusters. To investigate the influence of this external field effect on our sample, we follow~\citet{bekenstein&milgrom84,famaey_etal07b}, and express this effect in terms of the modification $\tilde{\nu}-$function. For the simple form, the $\tilde{\nu}-$function becomes
\begin{equation}
    \tilde{\nu}(x_N)= {\frac{1}{2}}\left(1-{\frac{g_e}{g_N}}+\sqrt{4(\frac{a_0}{g_N}+\frac{g_e}{g_N})+(1-\frac{g_e}{g_N})^2}\right),
	\label{eq:nu_simple_e}
\end{equation}
where $g_e$ denotes the external field. In Fig.~\ref{fig3}, we show how the external field effect will affect the slope of the MOND MP by considering $g_e=0.1 a_0$, $0.5 a_0$, and $a_0$. Although overall the external effect has only small impact on the best-fitting coefficients of the MOND MPs in our sample, we do find that a stronger external field will move the plane a little bit closer to the Newtonian one. In reality, however, not all galaxies in the sample will be affected by an external field as strong as $g_e=a_0$. To consider a more accurate case, we thus roughly model the external field based on the Fig.~\ref{fig3} of~\citet{richtler_etal11}, and classify the 16 galaxies into three groups. We assign $g_e=a_0$ to the two galaxies (GMP $3414$ and $3510$) that are located at the inner part of the cluster, assign $g_e=0.5 a_0$ to GMP $3792$, $2417$, and $2440$, whose projected distances from the center of the cluster are between $0.56$ and $0.67$ Mpc, and assign $g_e=0.1 a_0$ to the other 11 galaxies which are at least 1.45 Mpc away from the center. The MOND MP of this more realistic modeling of the external field effect is consistent with the theoretical prediction of Eq.~(\ref{eq:MMP}).

Finally, to complete our analysis, we also consider the Jaffe model as an alternative to the mass profile in our analysis (lower panel of Fig.~\ref{fig3}). Although the influence of the external field effect on the slope of the MOND MP is a little bit stronger in the Jaffe model than that in the Hernquist model, the MOND MP based on our modeling of the external field still agrees well with the theoretical prediction.
In general, we do not find any obvious disparity between the results of the Hernquist and the Jaffe models.

\section{Discussion and conclusion}\label{V}
This paper, for the first time since the existence of the FP was confirmed in 1987, successfully explain the origin of the FP without any fine-tuning.
We explicitly show that the tilt of the fundamental plane can be naturally explained in the framework of MOND by a combined effect of anisotropy of velocity dispersion and modification of gravity. All the parameters involved in our model are either well-motivated, such as the anisotropy profile $\beta_e$~\citep{vanAlbada82,sanders00}, and a universal stellar $M/L$~\citep{kroupa01}, or consistent with other independent observations~\citep{famaey&binney05,zhao&famaey06,famaey_etal07}, such as the simple form of the interpolating function. This is striking, because although the MONDian interpolating function might arguably be contrived to replace the requirement of dark matter in spiral galaxies, it is never the case for the slope of the FP. Everything just fits together naturally.

 We shall emphasize that the success of $\Upsilon_{\rm{Krou}}$ or $\Upsilon_{\rm{Salp}}$ at removing the tilt of the observed FP does not mean that our results rebut a non-universal IMF in MOND. Indeed, under the same conditions, $\Upsilon_{*,\rm{dyn}}$, a non-universal $M/L$ in~\citet{Thomas_etal11}, will also match the empirical MOND MP (but not the MFP) with its theoretical counterpart. Although $\Upsilon_{*,\rm{dyn}}$ contains the contribution of dark matter and only provides an upper limit to stellar mass, this result does support that a slight variant of stellar $M/L$ among galaxies is not ruled out. Hence, our finding does not necessarily contradict with~\citet{sanders00}, but disagrees with~\citet{tortora_etal14} at that MOND must require variant IMF to fit observations.

Over the past years, much attention has been paid to elliptical galaxies in MOND~\citep{Cardone_etal11,Chiu_etal11,Ferreras_etal12,Milgrom12,chae&gong15,Tian&Ko16}, partly because it is important to learn whether or not MOND is as successful in elliptical galaxies as in spiral ones. Unlike cluster of galaxies, where massive neutrino might accumulate and contribute to invisible matter~\citep{sanders07,Milgrom08}, any requirement for dark matter in galaxies will be devastating to the MONDian paradigm. This paper offers an independent piece of evidence to support the success of MOND, either as a fundamental theory or some effective theory, at galactic scales.

%Of various phenomenon belong to elliptical galaxies, how to understand the existence of the observed FP and its tilt with respect to the virial plane is still prevailing. In this paper, we explicitly show that this tilt can be naturally explained in the framework of MOND if we assume a universal $\Upsilon$, anisotropy of elliptical galaxies, and the simple form of interpolating function, all of which are in accordance with our understandings of elliptical galaxies and MOND. Indeed, our findings are consistent with Ref., and directly answer the concerns about the existence of a viable MFP in the introduction of Refs.

\section*{Acknowledgements}
We would like to thank Mordehai Milgrom and David Koo for helpful comments. MCC would also like to thank CMK for the hospitality during his stay in NCU.
CGS and MCC are partly supported by the Chinese National Nature Science Foundation No. 11433003, the Shanghai Science Foundation No. 13JC14044400 and National Key Project 973 No. 2014CB845704. CMK is supported in part by the Taiwan Ministry of Science and Technology grants MOST 104-2923-M-008-001-MY3 and MOST 105-2112-M-008-011-MY3.

%%%%%%%%%%%%%%%%%%%%%%%%%%%%%%%%%%%%%%%%%%%%%%%%%%

\bibliographystyle{apsrev4-1}
\bibliography{bib_list_prd} % if your bibtex file is called example.bib

\end{document}